# Application of the "Among" Leadership Model to Improve Teacher Work Discipline

Rusman [1] , Sunarti [1] , Raditya Bayu Rahadian [2]
[1] Education and Culture Office of South Bangka, Islands of Bangka Belitung, Indonesia
[2] Education Management, Graduate Schools, Yogyakarta State University, Indonesia

**Abstract:-** This study aims to improve teacher work discipline related to their duties as classroom teachers. This research is a school action research conducted for four months starting in October 2018 to January 2019 at SDN 11 Simpang Rimba, South Bangka, Indonesia in the 2018/2019 academic year. The data collection instrument used observation sheets and documentation. Data analysis used quantitative data analysis techniques with descriptive statistics. The results of this study indicate that the application of the "Among" leadership model can improve teacher work discipline in the aspects of arriving on time, work hours fulfillment, and prepare lesson plans.

*Keywords: "Among" Leadership Model, Teacher Work Discipline.*

## I. INTRODUCTION

The formation of good attitudes and behavior is strongly influenced by the maturity of discipline. Discipline is reflection of high motivation which will generate a passion for work which if maintained continuously will form a high dedication to their work. Teachers who have high dedication to their work will bring out a sense of love for work which is manifested in learning activities both indoors and outdoors. He will realize the consequences of a teacher who will be an example in every word, attitude, and deed, both at school as a teacher and outside of school when interacting as part of society.

Good discipline in schools is essential to ensure that all pupils can benefit from the opportunities provided by education (Parsons, 2018:530). A disciplined teacher is a forerunner to the formation of students with character and achievement not only in the school environment but also in the community. Starting from before leaving for school, being at school, going home, until returning to school is a series of deep love for the profession. Because of this, he will only show goodness and benefit to his students and to his fellow teachers as a personal mirror of a teacher.

Teachers who are motivated can be seen from their discipline in carrying out their work. Discipline towards work time, the fulfillment of working hours, and discipline in preparing for learning are forms of attitudes and concrete actions for teacher work discipline (Rosdiana, 2018:102). In the long term, this discipline will shape the character of the school which will be embedded into the school culture so that it becomes the superior value of the school.

The situation at SDN 11 Simpang Rimba is still far from the ideal criteria expected. The number of teachers who meet the criteria for good work discipline is still small. Teacher delays in coming to school, teachers have not met working hours, and do not prepare lesson plans are still common. Based on observations on indicators of teacher work discipline, it is concluded that the work discipline of teachers at SDN 11 Simpang Rimba is still low. It can be seen from the average score of the teacher arriving on time is 2.63 (medium), the work hours fulfillment score is 2.06 (low), and the score for preparing lesson plans is 1.56 (very low).

The reflections carried out found several causes for the low work discipline of teachers, including non-compliance with rules, leaders not giving good examples, and lack of appreciation and reinforcement from leaders. If this condition drags on, it will cause other bigger problems so that a solution must be found immediately.

One of the ways to develop teacher work discipline is by applying the right leadership model in certain situations and conditions as a value that underlies the formation of teacher work discipline. As stated by Patrick Duignan that leaders must be able to find a balance between achieving organizational goals with individual goals in them. Therefore, leaders need an ethical and values-based framework to make good decisions in such situations (Duignan, 2006). In this case, the leader needs to apply the right leadership model to discipline teachers in doing their job. Leaders must be able to provide good values, examples/role models, motivation, as well as efforts to enforce the rules that have been agreed upon.

One of the values-laden leadership models that can be applied is the "Among" leadership model. This model is extracted from the thoughts of Ki Hadjar Dewantoro which aims to foster work discipline that comes from within the teacher's person so that there is no feeling of compulsion and pressure in efforts to discipline teacher work (Hasanati, 2012:65). The "Among" leadership model is based on the principles of kinship, nature, and independence (Rahma & Setiadi, 2016:103). The principle of kinship is in accordance with the school environment, which is a unitary institution that has the same goal of developing the potential of its students. The natural principle of nature is in line with human nature which has the awareness to change towards goodness. The principle of independence is in accordance with the characteristics of teachers as adults who are free to determine their opinions and beliefs in the corridors of appropriate rules.

Based on the problems that have been described, the action hypothesis proposed in this research: The "Among" leadership model can improve teacher work discipline at SDN 11 Simpang Rimba, South Bangka, Indonesia.

## II. LITERATURE REVIEW

Leadership is the ability to influence a group of members to work towards goals and objectives. The source of influence can be obtained formally, namely by holding a managerial position in an organization (Hidayat & Machali, 2012:75). Another definition states that leadership is the ability to mobilize, motivate and influence people to want to take action directed to achieve goals regarding the success that is carried out, regarding the courage to make decisions about the activities carried out (Shulhan, 2013:9).

Leadership has moved from being leader centered, individualistic, hierarchical, focused on universal characteristics, and emphasizing power over others to a vision in which leadership is process centered, collective, context bound, nonhierarchical, and focused on mutual power and influence processes (Kezar & Carducci, 2009:2). This can require a big paradigm shift for many teachers and school administrators, since historically schools have relied on the use of reward and punishment. Not only are these reward and punishment systems ineffective in the long term, they are exhausting and time consuming. (Nelsen & Gfroerer, 2017:12).

Leader as a manager needs to lead the people that he or she is in charge of guiding toward a specific goal. This can include telling them what to do and when to do it, organizing the structure of the team members to highlight specific skills that each possesses, and even offering rewards for a job well done (Newton, 2016:11). The key challenges for educational leaders, especially principals, involved complex and often conflicting human relationships and interactions (Duignan, 2006:43).

The "Among" leadership model offers a leadership concept that focuses on the emergence of an awareness of responsibility for personal and organizational tasks through the role model of a leader towards that responsibility. The "Among" leadership model was initiated by Ki Hadjar Dewantoro, a figure of the Indonesian independence movement and a pioneer of Indonesian education who is also known as the Father of Indonesian Education (Yanuarti, 2017:240). This model is in line with transformational leadership theory which focuses on efforts to raise self-awareness for the people they lead about organizational behavior and goals. The awareness gained through one's own efforts will last longer and be steady so it is not easy to change (Hasanati, 2012).

The main principle of "Among" leadership: 1) Ing ngarso sung tulodho – a leader must be able, through his attitudes and actions, to make himself a model for the people he leads; 2) Ing madyo mangun karso - a leader must be able to arouse a spirit of initiative and creation in the people he leads; 3) Tut wuri handayani - a leader must be able to encourage its members to dare to walk in the front and be responsible (Hasanati, 2012:65).

Good discipline in schools is essential to ensure that all pupils can benefit from the opportunities provided by education (Parsons, 2018:530). High work discipline can arouse student motivation in learning. Teacher work discipline is closely related to compliance in implementing school rules, such as arriving on time, actively entering school, not leaving class before the lesson ends, and delivering learning material according to the lesson plan that has been prepared (Jumriah, Akib, & Darwis, 2016; Suwandi & Sajari, 2009).

Attitudes and actions of leaders that reflect the leadership model "Among": 1) Role model - The principal comes early and greets the teacher and students until the last one arrives; 2) Providing guidance and motivation - The principal ensures that the working hours are different from the required teaching hours. The working hours are greater than the required teaching hours to provide time for the teacher to evaluate and plan the learning that will be carried out; 3) Visits and personal dialogue - The principal visits the teacher who has arrived late, does not fulfill work obligations and does not make lesson plans.

## III. RESEARCH METHODOLOGY

This research is a school action research in order to improve teacher work discipline. The research method that will be used to answer the action hypothesis is the experiment which is carried out repeatedly in three cycles. As stated by Sugiyono (2015:133), action research is a type of experimentation in real-life situations. What is being experimented on or tried out is a plan of action or a hypothesis of action.

➢ *Research Design*
Testing the action hypothesis using pre-experimental design (non-designs) in the form of One-Group Pretest-Posttest Design to compare conditions before and after being given treatment/action. With this form, the results of action can be known accurately (Sugiyono, 2015:137-138). The pretest-posttest one group experimental design can be described as follows:

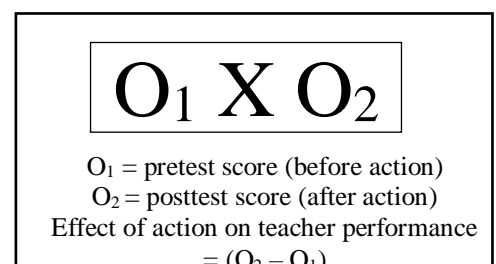

$O_1$ = pretest score (before action)
$O_2$ = posttest score (after action)
Effect of action on teacher performance
$= (O_2 - O_1)$

Fig. 1:- One Group Pretest-Posttest Experiment Design

The treatment of this research is the application of the "Among" leadership model. The experiment is "before-after", which compares the work discipline of the teacher before and after the application of the leadership model "Among". Each cycle of testing the action hypothesis is carried out in four processes, as stated by Cooper and Bedford that action research is an approach widely used to bring about school change. It is a practice involving reflection and action directed at transforming school practices and structures. The action research process has cycles consisting of planning, action, observation, and reflection phases that could involve an individual teacher or a whole school collective effort (Cooper & Bedford, 2017:267).

➢ *Population*
This research involved class teachers of SDN 11 Simpang Rimba, South Bangka, Indonesia in the 2018/2019 academic year. The number of classroom teachers who were the subject of this study amounted to 16 people, consisting of 10 female teachers and 6 male teachers.

➢ *Instrument*
The instruments used to observe changes in teacher work discipline are as follows. The score range refers to the five Likert scale: 5 = very good; 4 = good; 3 = medium; 2 = bad; 1 = very bad. In action research, the Likert scale is used to develop instruments to measure the attitudes, perceptions, and opinions of a person or group of people on the potential and problems of an object, action planning, and action results (Sugiyono, 2015: 199).

The work discipline of teachers in this study is described from the following indicators:

- Arriving on time
If the average monthly teacher working days are 25 days, then the Likert scale can be converted to score teacher attendance on time, as follows:
- delay within a month up to 3 times: 5
- delay within a month between 4 - 7 times: 4
- delay within a month between 8 - 12 times: 3
- delay within a month between 13 - 17 times: 2
- delay within a month more than 18 times: 1

- Work hours fulfillment
If the average working hours per week is 37.5 hours, then the Likert scale can be converted to score the level of fulfillment of teacher working hours, as follows:
- more than 32 hours a week: 5
- between 26 to 32 hours a week: 4
- between 20 to 25 hours a week: 3
- between 13 to 19 hours a week: 2
- less than 13 hours of a week: 1

- Prepare lesson plans
The lesson plan becomes the teacher's reference for carrying out learning activities. If all the learning that will be carried out is made a lesson plan and given a maximum value of 100%, then using the Likert scale can be converted the level of the lesson plan preparation, as follows:
- more than 84.4% lesson plans are ready to use: 5
- more than 70% to 84.4% lesson plans are ready: 4
- more than 51.8% to 70% lesson plans are ready: 3
- between 36.1% to 51.8% lesson plans are ready: 2
- less than 36.1% lesson plans are ready to use: 1

➢ *Data Collection*
The data collection techniques used in each cycle were observation and documentation. Observations were made every day on attendance, completeness of learning plan documents, and fulfillment of working hours after the application of the "Among" leadership model. Documentation is used to determine the level of teacher work discipline before and after the action.

➢ *Data Analysis*
The data analysis technique to find the effect of an action is a quantitative analysis using descriptive statistics to test the significance of the difference between O1 and O2 because the sample was not taken randomly. As stated by Sugiyono (2019:241), descriptive statistics are used to analyze data by describing or describing the collected data as it is. This includes the presentation of data through tables, graphs, diagrams, central tendency, and percentages. Research conducted on the population will clearly use descriptive statistics.

The action hypothesis testing in this research used descriptive statistical analysis through the comparison of the average teacher work discipline before the application and after the application of the "Among" leadership model. The steps in quantitative data analysis are scaling obtained from the assessment instrument using a Likert scale. To determine the level of teacher work discipline using scores: very good = 5, good = 4, medium = 3, bad = 2, very bad = 1 (Best & Kahn, 2006:331).

Data are presented in tabular form or frequency distribution. With this analysis, it will be seen that the tendency of teacher work discipline is shown to reach the degree: very low, low, medium, high, or very high. The five scale value conversion table can be seen in table 1:

| Degree | Score Interval | |
|---|---|---|
| Very High (VH) | $X > \overline{X_i} + 1{,}80\ SBi$ | $X > 4{,}21$ |
| High (H) | $\overline{X_i} + 0{,}60\ SBi < X \leq \overline{X_i} + 1{,}80\ SBi$ | $3{,}4 < X \leq 4{,}21$ |
| Medium (M) | $\overline{X_i} - 0{,}60\ SBi < X \leq \overline{X_i} + 0{,}60\ SBi$ | $2{,}59 < X \leq 3{,}4$ |
| Low (L) | $\overline{X_i} - 1{,}80\ SBi < X \leq \overline{X_i} - 0{,}60\ SBi$ | $1{,}79 < X \leq 2{,}59$ |
| Very Low (VL) | $X \leq \overline{X_i} - 1{,}80\ SBi$ | $X \leq 1{,}79$ |

Table 1. Likert Scale Conversion

From the assessment criteria shown in Table 1, it is obtained the standard level of teacher work discipline with the following details:
- Very High (VH) teacher work discipline if the average score obtained is greater than 4.21.
- High (H) teacher work discipline if the average score obtained is greater than 3.4 to 4.21.

- Medium (M) teacher work discipline if the average score obtained is greater than 2.59 to 3.4.
- Low (L) teacher work discipline if the average score obtained is greater than 1.79 to 2.59.
- Very Low teacher work discipline if the average score obtained is less than 1.79.

## IV. RESULT

Based on the results of the research that has been done, it can be described the profile of the score of teacher work discipline before the application of the leadership model among and after its application in the following table:

| Number of cycles | Teacher Work Discipline | | | |
|---|---|---|---|---|
| | Arriving on time | Work hours fulfillment | Preparing lesson plans | Mean |
| 1 | 3.31 | 3.13 | 2.69 | 3.04 |
| 2 | 4.00 | 3.50 | 3.44 | 3.65 |
| 3 | 4.44 | 4.19 | 4.13 | 4.25 |
| Mean | 3.92 | 3.60 | 3.42 | 3.65 |
| Before action | 2.63 | 2.06 | 1.56 | 2.08 |

Table 2:- Summary of Teacher Work Discipline Score's

Table 2 provides information that the mean achievement of teacher work discipline in the aspect of arriving on time before the action is 2.63 (low). When observed after the application of the "Among" leadership model, the mean score of the aspect of arriving on time in the first cycle increased to 3.31 (medium), increased again in the second cycle to 4.00 (high), and again increased in the third cycle to 4.44. (very high). This means that the discipline of teacher work in the aspect of arriving on time has an increasing trend. As shown in the chart 1.

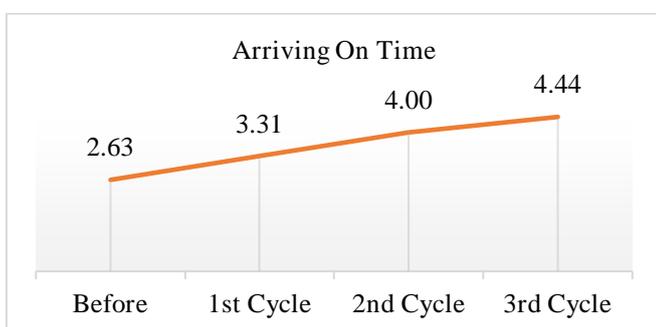

Chart 1:- Teacher Work Discipline Score's in the Aspects of Arriving on Time

Based on chart 1, it can be seen that the increase in teacher work discipline is consistent in the aspect of arriving on time after being given the action. This shows that the application of the "Among" leadership model can provide awareness to teachers to be more disciplined and maintain consistency to arrive on time to school.

Table 2 also provides information that the mean achievement of teacher work discipline in the aspect of working hours fulfillment before the action is 2.06 (low). When observed after the application of the "Among" leadership model, the mean aspect score of working hours fulfillment in the first cycle increased to 3.13 (medium), increased again in the second cycle to 3.50 (high), and increased again in the third cycle to 4.19 (high). This means that teacher work discipline in the aspect of working hours fulfillment tends to increase continuously, as shown in the chart 2.

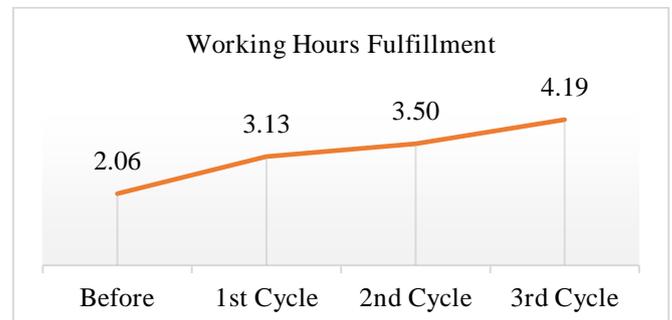

Chart 2:- Teacher Work Discipline Score's in the Aspects of Work Hours Fulfillment

Chart 2 shows the consistency of increasing teacher work discipline in the aspect of working hours fulfillment after being given action. This shows that the application of the "Among" leadership model can encourage teachers to be disciplined to make the best use of working hours so that the increase that occurs due to self-awareness can last longer and be sustainable.

Table 2 also provides information on the achievement of teacher work discipline in the aspects of prepare lesson plans. The score reached before the action was 1.56 (very low). After the application of the "Among" leadership model, the mean score achieved in the first cycle rose to 2.69 (medium), increased again in the second cycle to 3.44 (high), and increased again in the third cycle to 4.13 (high). This means that the discipline of teacher work in the aspect of prepare lesson plans tends to continue to increase, as shown in chart 3.

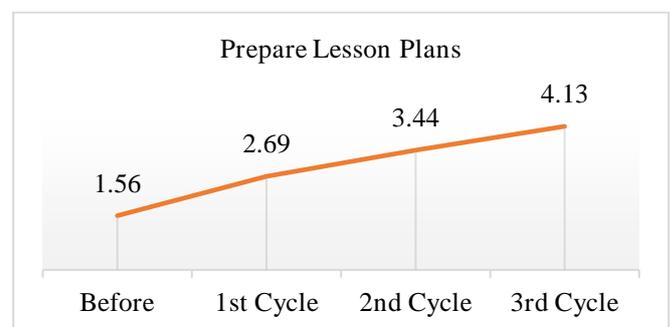

Chart 3:- Teacher Work Discipline Score's in the Aspects of Prepare Lesson Plans

Chart 3 shows the consistency of increasing teacher work discipline in the aspects of prepare lesson plans after being given action. This shows that the application of the "Among" leadership model can provide awareness and enthusiasm for teachers to be disciplined to prepare lesson plans as a guide in the implementation of learning.

The average achievement of teacher work discipline as a whole has also increased, as shown in chart 4.

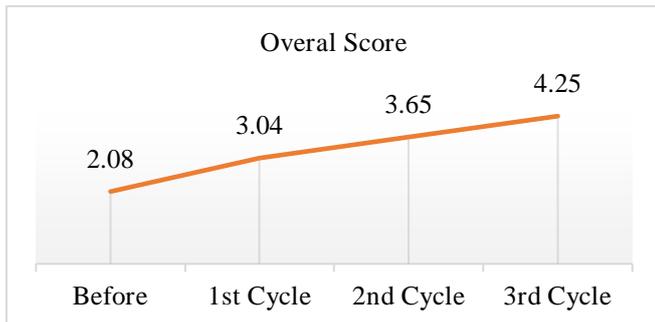

Chart 4:- Overall Teacher Work Discipline Score

Chart 4 shows that the average teacher work discipline has increased consistently. If all aspects of teacher work discipline are averaged, then a description of information on the level of teacher work discipline before and after the action can be obtained. Before the action, the mean of teacher work discipline was 2.08 (low), increased to 3.04 (moderate) in the first cycle, increased again to 3.65 (high) in the second cycle, and increased again in the third cycle to 4.25 (very high).

## V.　DISCUSSION

Good discipline in schools is essential to ensure that all pupils can benefit from the opportunities provided by education (Parsons, 2018:530). High work discipline can arouse student motivation in learning. Teacher work discipline is closely related to compliance in implementing school rules, such as arriving on time, actively entering school, not leaving class before the lesson ends, and delivering learning material according to the lesson plan that has been prepared (Jumriah et al., 2016; Suwandi & Sajari, 2009).

Strengthening teacher work discipline cannot be separated from the role of the leader, in this case, the principal. The leadership model that is in line with the efforts to generate teacher motivation regarding the awareness of work discipline must be applied, one of which is the "Among" leadership model. The leadership model initiated by Ki Hadjar Dewantoro is in line with transformational leadership theory which focuses on efforts to raise self-awareness for the people they lead about organizational behavior and goals. The awareness gained through one's own efforts will last longer and not easily changed.

The "Among" leadership model is based on the principles of kinship, nature, and independence (Rahma & Setiadi, 2016:103). The principle of kinship is in accordance with the school environment, which is a unitary institution that has the same goal of developing the potential of its students. The natural principle of nature is in line with human nature which has the awareness to change towards goodness.

The principle of independence is in accordance with the characteristics of teachers as adults who are free to determine their opinions and beliefs in the corridors of appropriate rules. The principle of independence that is upheld certainly does not necessarily negate the role of the principal in carrying out continuous monitoring and evaluation. Even so, the monitoring and evaluation that is carried out should be packaged with the principle of exemplary value and avoiding the principle of punishment, so that teachers feel appreciated and raise awareness.

In his leadership practice, the principal conducts discussions with all employees to formulate rules that must be agreed upon regarding the enforcement of work discipline. A discussion of the rules is carried out jointly so that employees feel involved in drafting and approving these rules. It is hoped that awareness of compliance with rules related to work discipline will be high.

Regarding the fulfillment of working hours, the principal emphasized that teachers must meet 37.5 hours per week. Absence and permission only when urgent. In fulfilling 37.5 working hours per week, teachers can take advantage of the time after teaching hours to evaluate and plan further learning activities, so that learning is better prepared to ensure student learning success. The teacher also has to compile all the lesson plans to guide the course of learning so that it will facilitate the achievement of learning objectives. The principal also reminded that lesson plans were also needed to fulfill teacher administrative duties which were worth credit points.

Efforts to enforce teacher work discipline are not based on punishment and threats, because it encourages more negativity, rejection, and rebellion. Therefore the principal chose to apply the leadership model among enforcing teacher work discipline. Actions taken are through modeling, providing guidance and motivation, as well as visits and personal dialogue. Principal give the example by arriving early and returning last unless attending invitations/external assignments that cannot be represented. Providing guidance and motivation in the form of ensuring that working hours are different from compulsory teaching hours. The working hours are greater than the required teaching hours to give the teacher time to evaluate and plan the learning that will be carried out. Visits and personal dialogues are carried out to visit and discuss problems experienced by teachers when they arrive late, do not meet working hours, and do not make lesson preparations. Visits are made to the home of the teacher concerned to avoid negative effects if the teacher is called to face during working hours. Besides, home visits serve as a friendship and closer emotional connection, so that teachers may be able to express feelings related to problems at home rather than at school.

Increasing teacher work discipline consistently shows that teacher adherence to work discipline arises because of self-awareness and encouragement so that its consistency can be maintained. It is very different if adherence to work discipline is influenced by threats, punishments, or things that come from outside the teacher himself, the consistency is difficult to maintain. This is evidenced by the achievement of the average teacher work discipline before the action which reached a score of 2.08 (low), increased to 3.04 (moderate) in the first cycle, increased again to 3.65 (high) in the second cycle, and increased again in the third cycle to 4.25 (very high). So it can be concluded that the hypothesis that Among" leadership model can improve teacher work discipline at SDN 11 Simpang Rimba, South Bangka Regency, Indonesia is proven empirical or acceptable.

## VI.　CONCLUSIONS AND SUGGESTIONS

Application of the "Among" leadership model can improve teacher work discipline in the aspects of arriving on time, work hours fulfillment, and prepare lesson plans at SDN 11 Simpang Rimba, South Bangka, Indonesia. The steps for implementing the "Among" leadership model for enforcing teacher work discipline are: a) formulating rules; b) discussion of joint regulations; c) agreement on collective rules; d) exemplary, guidance and motivation, personal visits; e) appreciation and reinforcement.

The "Among" leadership model can be applied in schools to improve teacher work discipline related to their duties. Even so, further research is needed to enrich other practical steps in implementing the leadership model among especially in schools.

## ACKNOWLEDGMENT

The authors are thankful to Education and Culture Office of South Bangka, Islands of Bangka Belitung, Indonesia, for supporting this research work.

## REFERENCES

[1]. Best, J. W., & Kahn, J. V. (2006). *Research in education* (10th ed., Vol. 28). Boston: Pearson. https://doi.org/10.2307/3345058
[2]. Cooper, R., & Bedford, T. (2017). Transformative Education for Gross National Happiness: A Teacher Action Research Project in Bhutan. In L. L. Rowell, J. M. Shosh, C. D. Bruce, & M. M. Riel (Eds.), *The Palgrave International Handbook of Action Research* (pp. 265–278). New York: Palgrave Macmillan.
[3]. Duignan, P. (2006). *Educational Leadership: Key Challenges and Ethical Tensions*. Cambridge: Cambridge University Press.
[4]. Hasanati, N. (2012). Alternatif model kepemimpinan pada era globalisasi. *Psikologika : Jurnal Pemikiran Dan Penelitian Psikologi*, *17*(1), 61–68. https://doi.org/10.20885/psikologika.vol17.iss1.art7
[5]. Hidayat, A., & Machali, I. (2012). *Pengelolaan Pendidikan : Konsep, Prinsip dan Aplikasi dalam Mengelola Sekolah dan Madrasah*. Yogyakarta: Kaukaba.
[6]. Jumriah, Akib, H., & Darwis, M. (2016). Disiplin Kerja Guru dalam Melaksanakan Tugas Pembelajaran di Sekolah Menengah Kejuruan Negeri 1 Barru. *Jurnal Office*, *2*, 156.
[7]. Kezar, A., & Carducci, R. (2009). Revolutionizing Leadership Development: Lessons from Research and Theory. In A. Kezar (Ed.), *Rethinking Leadership Practices in a Complex, Multicultural, and Global Environment: New Concepts and Models for Higher Education*. Virginia: Stylus Publishing.
[8]. Nelsen, J., & Gfroerer, K. (2017). *Positive Discipline Tools for Teachers: Effective Classrom Management for Social, Emotional, and Academic Success*. New York: Harmony.
[9]. Newton, P. (2016). *Leadership Models : Leadership skills*. bookboon.com The eBook company. Retrieved from http://linkinghub.elsevier.com/retrieve/pii/B9780750669016500069
[10]. Parsons, C. (2018). Looking for Strategic Alternatives to School Exclusion. In J. Deakin, E. Taylor, & A. Kupchik (Eds.), *The Palgrave International Handbook of School Discipline, Surveillance, and Social Control* (pp. 529–552). Cham: Palgrave Macmillan.
[11]. Rahma, A., & Setiadi, B. N. (2016). Gambaran pendidikan kepemimpinan melalui metode " among " di perguruan tamansiswa. *Jurnal Psiko-Edukasi*, *14*, 101–112.
[12]. Rosdiana. (2018). Meningkatkan kedisiplinan guru dalam melaksanakan tugas melalui penerapan reward di sd negeri 050745 pangkalan berandan tahun ajaran 2016/2017. *Jurnal Tabularasa PPS Unimed*, *15*(1). https://doi.org/10.1017/CBO9781107415324.004
[13]. Shulhan, M. (2013). *Model Kepemimpinan Kepala Madrasah dalam Meningkatkan Kinerja Guru*. Yogyakarta: Teras.
[14]. Sugiyono. (2015). *Metode penelitian tindakan komprehensif*. Bandung: Alfabeta.
[15]. Sugiyono. (2019). *Metode penelitian pendidikan*. Bandung: Alfabeta.
[16]. Suwandi, & Sajari. (2009). *Memahami Penelitian Kualitatif*. Jakarta: Rineka Cipta.
[17]. Yanuarti, E. (2017). Pemikiran Pendidikan Ki. Hajar Dewantara dan Relevansinya Dengan Kurikulum 13. *Jurnal Penelitian*, *11*(2), 237–266. https://doi.org/10.21043/jupe.v11i2.3489